\documentstyle[12pt,amssym,aasms4]{article}

\def \rsun {\ifmmode$R$_{\odot}\else R$_{\odot}$\fi}

\def \hcm {\hbox {\ifmmode $ H atoms cm$^{-2}\else H atoms cm$^{-2}$\fi}}
\def\approxgt{\mathrel{\hbox{\rlap{\lower.55ex \hbox {$\sim$}}
        \kern-.3em \raise.4ex \hbox{$>$}}}}
\def\approxlt{\mathrel{\hbox{\rlap{\lower.55ex \hbox {$\sim$}}
        \kern-.3em \raise.4ex \hbox{$<$}}}}

\newcommand {\sax} {{\it BeppoSAX }}

\lefthead{Malizia et al.}
\righthead{Hard X-ray detection of the high redshift quasar 4C 71.07}

\begin{document}

\title{Hard X-ray detection of the high redshift quasar 4C 71.07}

%   \subtitle{}
\author{A. Malizia\altaffilmark{1,2}, L. Bassani\altaffilmark{3},
A.~J. Dean\altaffilmark{1}, M. McCollough\altaffilmark{4},
J. B. Stephen\altaffilmark{3}, S.~N. Zhang\altaffilmark{5,6},
W.~S. Paciesas\altaffilmark{5}}

\altaffiltext{1}{Department of Physics and Astronomy, University of
Southampton,
SO17  1BJ, England}
\altaffiltext{2}{\sax Science Data Center, ASI, Via Corcolle, 19 I-00131
Roma , Italy}
\altaffiltext{3}{Istituto TESRE, CNR, Via Gobetti 101, I-40129 Bologna,
Italy}
\altaffiltext{4}{University Space Research Association, Huntsville, AL
35806, USA}
\altaffiltext{5}{University of Alabama in Huntsville, Huntsville, AL 35899}
\altaffiltext{6}{Space Sciences Lab., NASA, Marshall Space Flight Center,
Huntsville, AL 35812}

\begin{abstract}

BATSE/OSSE observations of the high redshift quasar 4C 71.07 indicate
that this is the brightest and furthest AGN so far detected above 20
keV.  BATSE Earth occultation data have been used to search for
emission from 4C 71.07 from nearly 3 years of observation. The mean
source flux over the whole period in the BATSE energy range 20--100
keV is (13.2 $\pm$1.06) $\times$ 10$^{-11}$ erg cm$^{-2}$ s$^{-1}$
corresponding to a luminosity of 2 $\times$ 10$^{48}$ erg s$^{-1}$.
The BATSE light curve over the 3 years of observations shows several
flare-like events, one of which (in January 1996) is associated with
an optical flare (R=16.1) but with a delay of 55 days. The OSSE/BATSE
spectral analysis indicates that the source is characterized by a flat
power spectrum ($\Gamma$$\sim$ 1.1- 1.3) when in a low state; this
spectral form is consistent within errors with the ASCA and ROSAT
spectra.  This means that the power law observed from 0.1 to 10 keV
extends up to at least 1 MeV but steepens soon after to meet EGRET
high energy data. BATSE data taken around the January 1996 flare
suggests that the spectrum could be steeper when the source is in a
bright state.  The $\nu$F$\nu$ representation of the source is typical
of a low frequency peaked/gamma-ray dominated blazar, with the
synchrotron peak in the mm-FIR band and the Compton peak in the MeV
band. The BATSE and OSSE spectral data seem to favour a model in which
the high energy flux is due to the sum of the synchrotron self-Compton
and the external Compton contributions; this is also supported by the
variability behaviour of the source.

\end{abstract}

\keywords{galaxies: active -- gamma rays: observations -- X-rays: galaxies }

\section{Introduction}

4C 71.07 (0836+710) is a high redshift (z=2.172) quasar characterized
by a flat radio spectrum ($\alpha$=--0.33, Kuhr et al. 1981), a 5 GHz
flux greater than 1Jy (Kuhr et al. 1981), and both stationary and
superluminally moving components in a bright one-sided jet emerging
from the core (Krichbaum et al. 1990, Hummel et al 1992). Although the
source has unusually strong polarization from 21 cm (6.8$\%$) down to
9mm (9.5$\%$, Krichbaum et al. 1990) it is classified as a low
polarization quasar (LPQ) as optically the polarization is only
1.1$\%$ (Impey $\&$ Tapia 1990).  It has been monitored since 1989 in
the optical band and found to display at least 2 flares in February
1992 (von Linde et al. 1993)) and November 1995 (Raiteri et al. 1998):
typically the maximum brightness in R magnitude is around 16.1-16.5
with a $\Delta$R $\le$ 1.3 from minimum to maximum. The 1992 flare may
have been detected also at mm and cm frequencies but with a delay of
0.1-0.5 years (Otterbein et al. 1998). At soft X-ray energies (0.1-2
keV) the source underwent a flux decrease by a factor of $\sim$ 2
between March and November 92 without any spectral change (Brunner et
al. 1994) implying a high flux level close to the optical flaring
period; the source was still dim when reobserved by ASCA in March 1995
(Cappi et al. 1997).  The X-ray spectrum is characterized by a flat
photon index ($\sim$ 1.5) from 0.1-10 keV and by intrinsic X-ray
absorption which is variable ($\Delta$ N$_{H}$ $\sim$ 8 $\times$
10$^{20}$ cm$^{-2}$) on a timescale of 0.8 years. The observed X-ray
flux is typically 1.4 $\times$ 10$^{-11}$ erg cm$^{-2}$ s$^{-1}$ in
the 2-10 keV band corresponding to a luminosity of 2.1 $\times$
10$^{47}$ erg s$^{-1}$ (assuming here and in the following H$_{o}$=50
Km s$^{-1}$ Mpc$^{-1}$, q$_{o}$=0).  4C71.07 has also been observed by
EGRET on different occasions but not always detected (Thompson et al
1993, Mukherjee et al. 1997): in particular soon after the optical
flare observed in 1992 the gamma-ray flux was also high.  VLBI
monitoring of the source has further indicated the ejection of a new
jet component shortly after the time of the gamma/X/optical/radio
outburst (Otterbein et al. 1998). The gamma-ray source spectrum is
among the steepest at these energies ($\alpha$=2.4, Thompson et al
1993).  It is quite a bright source in the gamma-ray domain with a
50-200 MeV flux of 1.5 $\times$ 10$^{-10}$ erg cm$^{-2}$ s$^{-1}$ and
an isotropic luminosity of $\sim$2.2 $\times$ 10$^{48}$ erg s$^{-1}$;
note however that the gamma radiation is likely to be beamed (Thompson
et al.  1993).  We report here on the BATSE and OSSE detection of this
source in the unexplored energy range from 20-500 keV: this is the
farthest AGN so far detected at these energies and probably the
brightest.

\section{Temporal behaviour from BATSE data}

4C 71.07 is one of the few high redshift objects observed with
instruments onboard CGRO. BATSE data used for this work were collected
by the Large Area Detectors (LADs) and analysed in Earth occultation
mode (Harmon et al. 1992).  Since the occultation technique reaches
its peak sensitivity below $\sim$ 140 keV, the search for emission
from 4C71.07 was carried out in the energy range 20-100 keV. BATSE
data from nearly 3 years of observations (December 1994- October 1997)
have been analysed to extract a mean flux and a source light curve
over this period. The median statistic was used to remove about 5$\%$
of outlier data points in order to reduce systematic errors. In the
case of this source no contamination from known nearby hard X-ray
sources was found and so further cleaning was not applied (see Malizia
et al. 1999 for details on BATSE data analysis).  Counts were
converted to fluxes by folding a power law model of photon index 1.5
(similar to that observed in X-rays) with the instrumental response
function. The estimated uncertainty introduced by an arbitrary value
of the photon index is of the order of 10-20$\%$ for
$\Delta$$\alpha$=$\pm$ 0.3 (Malizia et al. 1999).

The mean source flux over the analysed period is (1.32 $\pm$0.11)
$\times$ 10$^{-10}$ erg cm$^{-2}$ s$^{-1}$ implying a detection
confidence level of 12$\sigma$; even increasing the error by 80$\%$ to
account for residual systematic uncertainties evaluated by examining
"blank sky fields" (Malizia et al. 1999), the detection remains at
7$\sigma$ level and so it is highly significant.  However to further
check the reality of the source, we have also produced an image of the
sky region around 4C71.07 using the occultation imaging method
described by Zhang et al. (1993).  Currently this method only allows a
few weeks of data to be summed and so 35 days of data from TJD 10099
to TJD 10134 around the maximum brightness, were used to create the
image showed in Figure 1.

\placefigure{fig1}

It is evident from Figure 1 that the source is detected also in this
limited period of time at a significance level of $\sim$3.5$\sigma$.
Having determined the uncertainty in the source position from Figure
1, we have further checked in SIMBAD and NED the existence in that sky
area (1$^{\circ}$ around the source position) of other potential high
energy sources. A number of objects detected by ROSAT are reported
within this region but their soft X-ray (0.24-2.0 keV) count rates,
typically in the range 0.002--0.05 counts/s, are well below that of
4C71.07 (0.4-0.7 counts/s in the same energy band).  None of the other
optical sources are candidates for high energy emission. We therefore
conclude that not only has BATSE detected a high energy source, but
that this source is the quasar 4C 71.07.  The long term light curve of
4C71.07 is shown in Figure 2, where BATSE data have been combined in
60 day bins to enhance the statistical significance; note that here
the errors considered are only statistical.  A few flare like events
are visible around April 1995 (TJD9750--9900), January 1996
(TJD10000--10200), November 1996 (TJD10350--10400) and May 1997
(TJD10500--10600) separated by periods of low flux when the source is
barely visible by BATSE.  The crosses in the Figure 2 correspond to
the optical (R band) measurements (Raiteri et al. 1998) which have
been superimposed to the BATSE data to better investigate possible
correlations.  In fact, the first event corresponds to a period of
medium brightness in optical (i.e. the flux in R was $\sim$ 20 $\%$
lower than observed during the 1992 flare (Von Linde et al. 1993) and
so compatible with the BATSE high state.  More interesting is the
second flare which corresponds to a period of optical flaring activity
monitored in the R band (Raiteri et al. 1998): the source was at its
historical maximum (R=16.1) on November 20, 1995 (TJD10042) about 55
days before the BATSE peak on January 14, 1996 (TJD10097). The optical
peak is however highly structured with a second peak on December 8,
1995 and a third on January 7, 1996; both peaks are characterized by a
lower intensity in R than the first one.  BATSE data do not have the
same statistical quality of the optical monitoring to resolve these
structures but the data are indicative of a similar broad structure.
Unfortunately no optical coverage is available during the other flare
like events detected by BATSE.  The source reached a 20-100 keV flux
of $\sim$3 $\times$ 10$^{-10}$ erg cm$^{-2}$ s$^{-1}$ during the peak
but was back to its mean flux about 3 months later; the upper limit on
the minimum timescale to double the flux is $\sim$ 60 days implying a
size for the emission region of $\le$ 5 $\times$ 10$^{16}$ cm.  Note
also that the change in flux observed during the November 1995/January
1996 event is a factor 4.5$\pm$0.5 in hard X-rays and 1.6 in optical.

The mean observed 20-100 keV flux translates to a luminosity of 2.6
$\times$ 10$^{48}$ erg s$^{-1}$ : this makes 4C 71.07 the brightest
and farthest AGN so far detected by BATSE.

\placefigure{fig2}

\section{Spectral behaviour: OSSE/BATSE  data}

4C 71.07 was observed continuously by OSSE in 1996 during the period
April 4 to 24 (TJD10178--10198). OSSE data reduction followed standard
procedures.  Energy spectra were accumulated in a sequence of two
minute measurements of the source field alternated with two-minute
measurements of background fields.  The data from each of the four
OSSE detectors were analysed individually and found compatible with
each other, then they were summed and the appropriate response matrix
was generated.  The source was detected at a significance level of
5.5$\sigma$.  The data have been rebinned into 8 broad channels from
50 to $\sim$600 keV and then fitted using the XSPEC package, version
10.0.  Due to the poor statistical quality of the data, only a simple
power law model can be used to fit the OSSE spectrum. The resulting
power law slope is flat ($\Gamma$ = 1.1$\pm$0.3, 90$\%$ confidence
level) and compatible within errors with the X-ray index measured by
ASCA and ROSAT. The data indicates that the power law observed from
0.1 to 10 keV extends at least up to 600 keV (or about 2 MeV in the
QSO rest frame) but steepens soon after to meet EGRET high energy
data.  The OSSE count rate spectrum and data to folded model ratio for
this simple power law, which gives a good description of the data
($\chi^{2}$ = 4.11 for 6 d.o.f.), is shown in Figure 3.

\placefigure{fig3}

The 50-600 keV flux is 2.5 $\times$ 10$^{-10}$ erg cm$^{-2}$ s$^{-1}$,
corresponding to a luminosity of $\sim$5 $\times$ 10$^{48}$ erg
s$^{-1}$.  
Inspection of Figure 2 indicates that at the time of the OSSE
observation the BATSE flux decreased from 21$\pm$5 to 7$\pm$5 $\times$
10$^{-11}$ erg cm$^{-2}$ s$^{-1}$, while the OSSE spectrum provides a
mean flux of 4.3 $\pm$ 1.1 $\times$ 10$^{-11}$ erg cm$^{-2}$ s$^{-1}$
over the same period.  Taking into account that the cross calibration
between these two instruments gives a BATSE flux normalization up to
40$\%$ higher than OSSE in the same energy band (Malizia et al. 1999),
we assume that the two data set are compatible within the respective
uncertainties and that during the OSSE observation the source was in a
lower state of activity than during the January 1996 flare.
Extrapolation of the OSSE spectrum to lower energies provides a 2-10
keV flux in the range 0.2-1.4 $\times$ 10$^{-11}$ erg cm$^{-2}$
s$^{-1}$, slightly lower than 1.4 $\times$ 10$^{-11}$ erg cm$^{-2}$
s$^{-1}$ reported by ASCA in 1995 but still consistent.  Comparison
with the ROSAT flux range in the same energy band (0.9-1.9 $\times$
10$^{-11}$ erg cm$^{-2}$ s$^{-1}$) further strengthens the idea that
4C71.07 was in a lower state of activity at the time of the OSSE
observation than during the flare.

BATSE spectra related to 60 days of data around the January 1996 peak
(TJD 10063-10123, data set A) and the OSSE observation
(TJD10178-10238, data set B) were also extracted and analysed over the
50-600 keV energy range.  The source was detected at a significance
level of 9.5 and 4 $\sigma$ respectively. The data were then fitted
again with a simple power law which in both cases gave a satisfactory
fit with $\chi^{2}_{\nu}$= 0.93 and 0.63 (12 degrees of freedom) and
photon indices of 2.3 $^{\rm+0.4}_{\rm-0.3}$ and 1.3
$^{\rm+0.5}_{\rm-0.4}$ for sets A and B respectively.  These two BATSE
measurements are reported in the broad band spectral energy
distribution of the source (Figure 4 in the next section).
Interestingly, set B has a spectral index similar to that obtained by
OSSE, although the intensity is higher as explained above. Set A
instead is best represented by a steeper photon index although only
marginally so.  This is in contrast with the typical {\it flatter when
brighter} behaviour of Blazars and therefore likely to be a key point
in source modelling. This is the second source in which such behaviour
is observed (see Ghisellini et al. 1999 for the case of PKS0528+134)
and so although the evidence is marginal it is however an important
observational constraint.

\section{Discussion}

Several models have been put forward to explain the overall broad band
spectral energy distribution (SED) of Blazars.  In the widely adopted
scenario, a single population of high energy electrons in a
relativistic jet radiate from the radio/Far-IR to the UV/soft X-ray by
the synchrotron process and at higher frequencies by inverse Compton
scattering soft-target photons present either in the jet (SSC model)
or in the surrounding ambient (EC model) or in both (Ghisellini et al.
1998 and references therein). In the following, by EC we mean a model
in which both the SSC and EC contributions to the high energy emission
are present, while the external photons are neglected in the SSC
model.  Then in the $\nu$F$\nu$ representation of Blazar SED, two
peaks corresponding to the synchrotron and inverse Compton components
should be evident.  The non simultaneous broad band spectral energy
distribution of 4C 71.07, adapted from Ghisellini et al. (1998), is
shown in Figure 4, with the inclusion of OSSE, BATSE and other data.
A broad hump associated with the Compton peak is strongly inferred by
the high energy data (OSSE/BATSE versus EGRET) and localized in the
10$^{20}$-10$^{22}$ Hz band; the corresponding synchrotron peak falls
in a region of the spectrum localized between 10$^{12}-10^{14}$ Hz,
where only upper limits from IRAS are available.  Nevertheless, the
data are consistent with the object being a low frequency peaked or
{\it red} blazar: the synchrotron and Compton spectra peak in the
mm-FIR (far-infrared) and MeV bands respectively and the X-gamma
radiation completely dominate the radiative output.  In these powerful
blazars, the contribution from the external radiation to the cooling
is supposed to be the greatest (Ghisellini et al. 1998); thus the EC
model should be favoured with respect to the SSC model.  In Figure 4
the SED from the SSC and EC models are superposed to the data of 4C
71.07; the model parameters are those used by Ghisellini et al. (1998)
and reported in their Table 2 of the Appendix. It is evident from the
figure that neither version of the theoretical modelling fits the data
well, since both models are unable to reproduce the radio data below
100 GHz; this is a problem often encountered in the interpretation 
of blazar SED and is generally solved by assuming that the extra model 
emission come from more extended source region. 
Furthermore, while the IRAS upper limits favour the SSC
model over the EC, the high energy data support the inverse. A major
drawback in such type of analysis, and also a possible explanation for
obtaining contradictory results, is that the fact the broad band data come
from different epochs: source variability, which is extreme in
4C71.07, may invalidate tests based on non contemporaneous
measurements.

\placefigure{fig4}

Nevertheless, we can rely on alternative ways to constrain models, for
example by examining the possible flatter state of the source when
brighter.  Ghisellini et al. (1999) have recently proposed a scenario
to explain this behavior in PKS 0528+134, a source very similar to
4C71.07 (i.e. a low frequency peaked/gamma-ray dominated blazar).
Their suggestion is that the high energy flux is produced by both the
SSC and EC processes. If the radiation densities of these two
components are comparable then the EC component dominates and is
entirely responsible for the gamma-ray emission. If, on the other
hand, the source power increases then the SSC component becomes more
important and therefore is the main contributor to the high energy
flux over a larger energy range. If this also happens in 4C71.07, then
the two gamma-ray states could correspond to different electron
injected powers (larger and flatter in the high state).  In any case,
a contribution from both the SSC and EC components is required in each
state.

Another method to distinguish between models is by studying the
variability behaviour of the source.  In the SSC model one expects
that a change in the electron spectrum causes a larger variability in
the inverse Compton emission than in the synchrotron emission; in
particular in a one-zone model (which should be the case here, since
the agreement between OSSE and ROSAT/ASCA data suggests that the
emission in the 0.002-1 MeV originates in the same region), the
inverse Compton peak flux should vary quadratically with the
synchrotron peak flux.  In the EC case, instead, the inverse Compton
emission should vary linearly for changes in the electron spectrum.
It is difficult to assess the source behaviour given the frequencies
(optical, on the descending part of the synchrotron hump and hard
X-rays, on the ascending part of the inverse Compton hump) available
for the flare monitoring.  Nevertheless, the data are in agreement
with the SSC model during the optical flare, i.e. the quadratic
variation condition was roughly satisfied (a factor of $\sim$ 4 in
hard X-rays versus a factor of $\sim$ 2 at optical frequencies).  Note
however that in the SSC scenario, light curves at different
frequencies are expected to be quasi-symmetric.
Instead, if the contribution of seed photons produced
external to the jet is relevant, the resulting light curves of the
synchrotron and inverse Compton emission can show considerable time
lags as, for example, if an active blob passes through the seed
photons region some time after its ejection (Chiaberge et al.  1998).
Unfortunately our coverage is too limited to test for any temporal
relationship between the optical and hard X-ray fluxes but we cannot
exclude that the flares observed at these frequencies correspond to
the same event with a time lag of approximately 55 days as also indicated by
Otterbein et al (1998) during the 1992 flare.  Thus the data being
compatible with both scenarios point again to a situation where both SSC and
EC components may contribute to the high energy emission.  Clearly
further monitoring of the source possibly at peak frequencies in
conjunction with VLBI observations to follow jet activity will be
extremely useful for a better understanding of the mechanism at work.

\section{Conclusions}

We report the result of the BATSE/OSSE observations of the sky region
containing the high redshift QSO 4C71.07. High energy emission,
attributed to the QSO, have been detected up to about 1 MeV; this
makes 4C71.07 the most distant AGN known to emit gamma-rays.  The
spectrum measured suggests that the power law observed at low energies
(0.1-10 keV) extends up to the MeV band; however the source spectrum
must steepen soon after to meet the high energy data measured by
EGRET. There is also evidence in our data for a steepening of the
spectrum during a high source state contrary to the usual "flatter
when brighter" behaviour of blazars.  BATSE monitoring of the source
in the 20-100 keV band over a 3 year period shows flare like events
one of which, in January 1996, is associated to an optical flare.
The source SED follows the general trend of
most luminous blazars showing two peaks, in the mid/far-infrared and
in the MeV band and a gamma-ray luminosity that dominates the overall
energy output.  Spectral data indicate that the high energy emission
is due to the sum of SSC and EC contributions with the former becoming
more important during the source high state. Hard X-ray monitoring of
the source during the 1996 optical flare also indicates the need for
both these components.

\acknowledgements 
We are grateful to Gabriele Ghisellini to provide us
the theoretical models used in this work to explain the data and
Claudia Raiteri for the optical measurements. We thank Tom Bridgman of
the Compton Observatory Science Support Center (GSFC) for retrieve the
OSSE data from public archive and Marco Malaspina for his support in
the implementation of the BATSE analysis package at the Istituto
Tecnologie e Studio delle Radiazioni Extraterrestri.

\clearpage

\smallskip
\def\ref{\par\noindent\hangindent 20pt}

\noindent
\section{References}
\vglue 0.2truecm

\ref{Brunner H., Lamer G., Worrall D.~M., Staubert R., 1994, A\&A, 287,436}
\ref{Cappi M., Matsuoka M., Comastri A., Brinkmann W., Elvis M.,
Palumbo, G.~G.~C., Vignali C., 1997, ApJ, 478, 492}
\ref{Chiaberge M., Celotti A. and Ghisellini G. 1998, astro-ph/9810335}
\ref{Ghisellini G., Celotti A., Fossati G., Maraschi L., Comastri A., 1998
MNRAS, 301, 451}
\ref{Ghisellini G., Costamante L., Tagliaferri G. et al., 1999, A\&A, in
press
(astro-ph/9906165)}
\ref{Harmon A. et al. 1992, The Compton Observatory Science Workshop,
NASA CP3137, eds C.~R. Shrader, N. Gehrels, B. Dennis, 69}
\ref{Hummel C.~A., Muxlow T.~W.~B., Krichbaum T.~P., Quirrenbach A.,
Schalinski C.~J., Witzel A., Johnston K.~J., 1992, A\&A, 266, 93}
\ref{Impey C.~D. and Tapia S., 1990, ApJ, 354, 124}
\ref{Krichbaum T.~P., Hummel C.~A., Quirrenbach A., Schalinski C.~J.,
Witzel A., Johnston K.~J., Muxlow  T.~W.~B, Qian S.~J, 1990 A\&A, 230, 271}
\ref{Kuhr H., Witzel A., Pauliny-Toth I.~I.~K., Nauber U., 1981, A\&AS, 45,
367}
\ref{Mcintosh D.~H., Rieke M.~J., Rix H.~W., Foltz C.~B., Weymann R.~J.,
     1999, ApJ, 514, 40}
\ref{Malizia A., Bassani L., Zhang S.~N., Dean A.~J., Paciesas, W.~S.,
Palumbo G.~G.~C., 1999, ApJ, 519, 637}
\ref{Mukherjee R., Bertsch D.~L., Bloom S.~D., et al. 1997, ApJ, 490, 116}
\ref{Otterbein K., Krichbaum T.~P., Kraus A., Lobanov A.~P., Witzel A.,
Wagner
S.~J., Zensus J.~A., 1998, A\&A, 334, 489}
\ref{Raiteri C.~M., Ghisellini G., Villata M., De Francesco G., Lanteri L.,
Chiaberge M., Peila A., Antico G., 1998, A\&AS, 127, 445}
\ref{Thompson D.~J., Bertsch D.~L., Dingus B.~L. et al., 1993, ApJ, 415L,
13}
\ref{von Linde J., Borgeest U., Schramm K.~J., Graser U., Heidt J., Hopp U.,
Meisenheimer K., Nieser L., Steinle H., Wagner S., 1993, A\&A, 267L, 23}
\ref{Zhang S.~N., Fishman G.~J., Harmon B.~A., Paciesas W.~S. 1993, Nature,
366, 245}
\ref{}

\clearpage

\figcaption[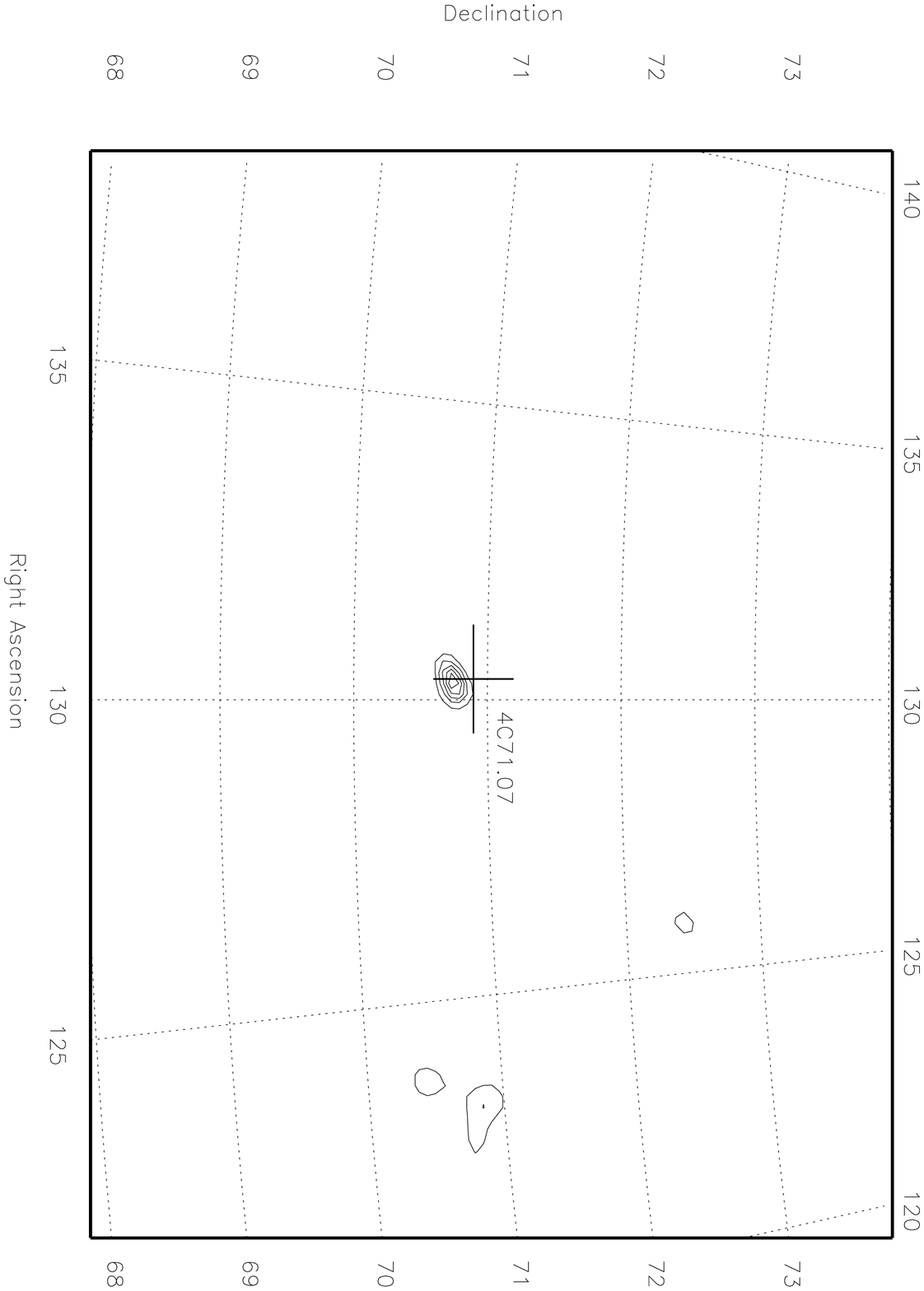]{BATSE image of the sky region around the QSO 4C 71.07
using 35 days of data (from TJD 10099 to TJD 10134). \label{fig1}}

\figcaption[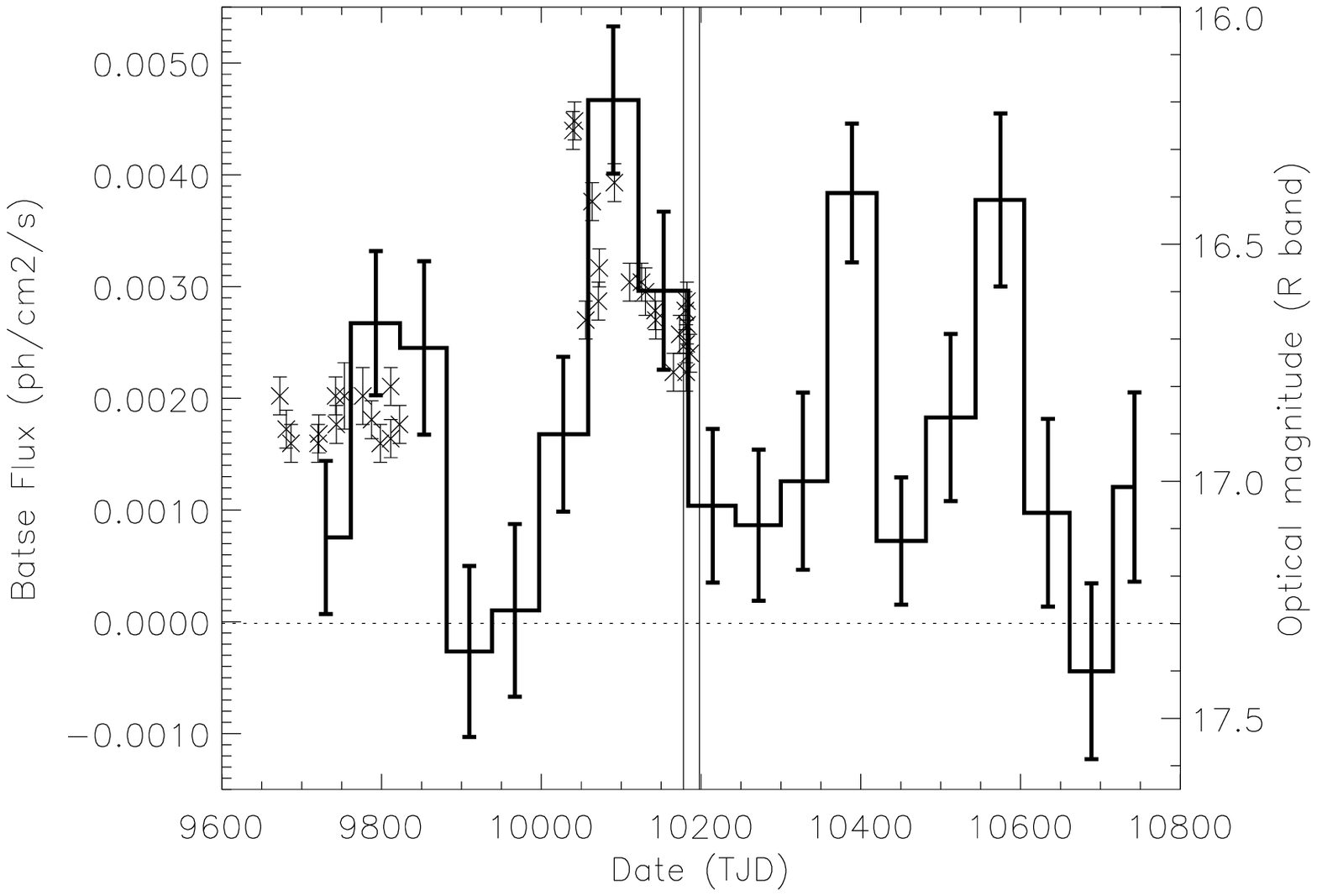]{BATSE light curve over 3 year period; each bin
corresponds to 60 days integration time.  The crosses superimposed to
the BATSE data correspond to optical (R band) measurements; the optical
minimum has been set to R-17.4 based on Von Linde et al.  (1993)
results.  The two vertical lines indicate the period of the OSSE
observation.
\label{fig2}}

\figcaption[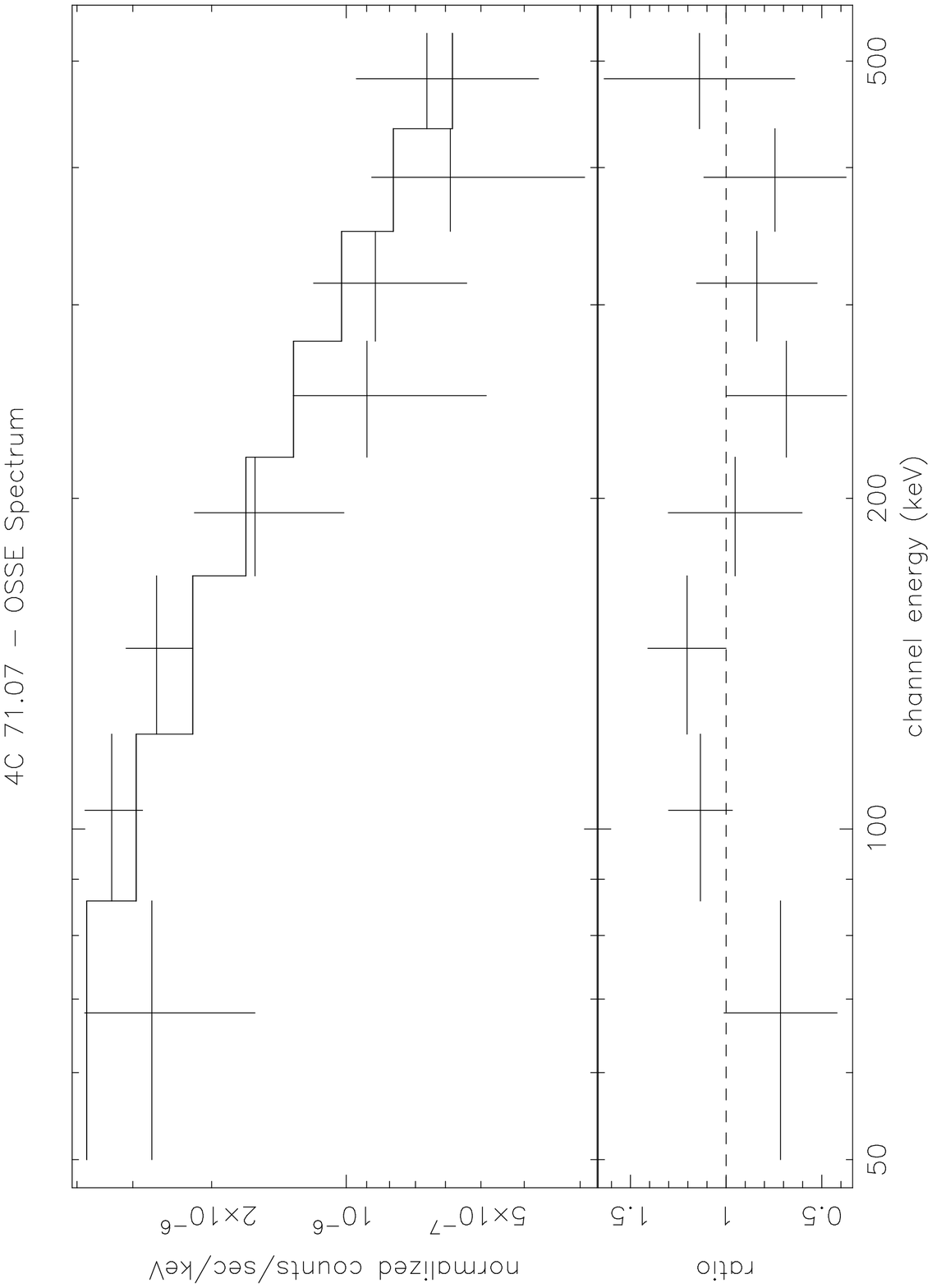]{OSSE count spectrum with residuals for a simple power
law model (see text). \label{fig3}}

\figcaption[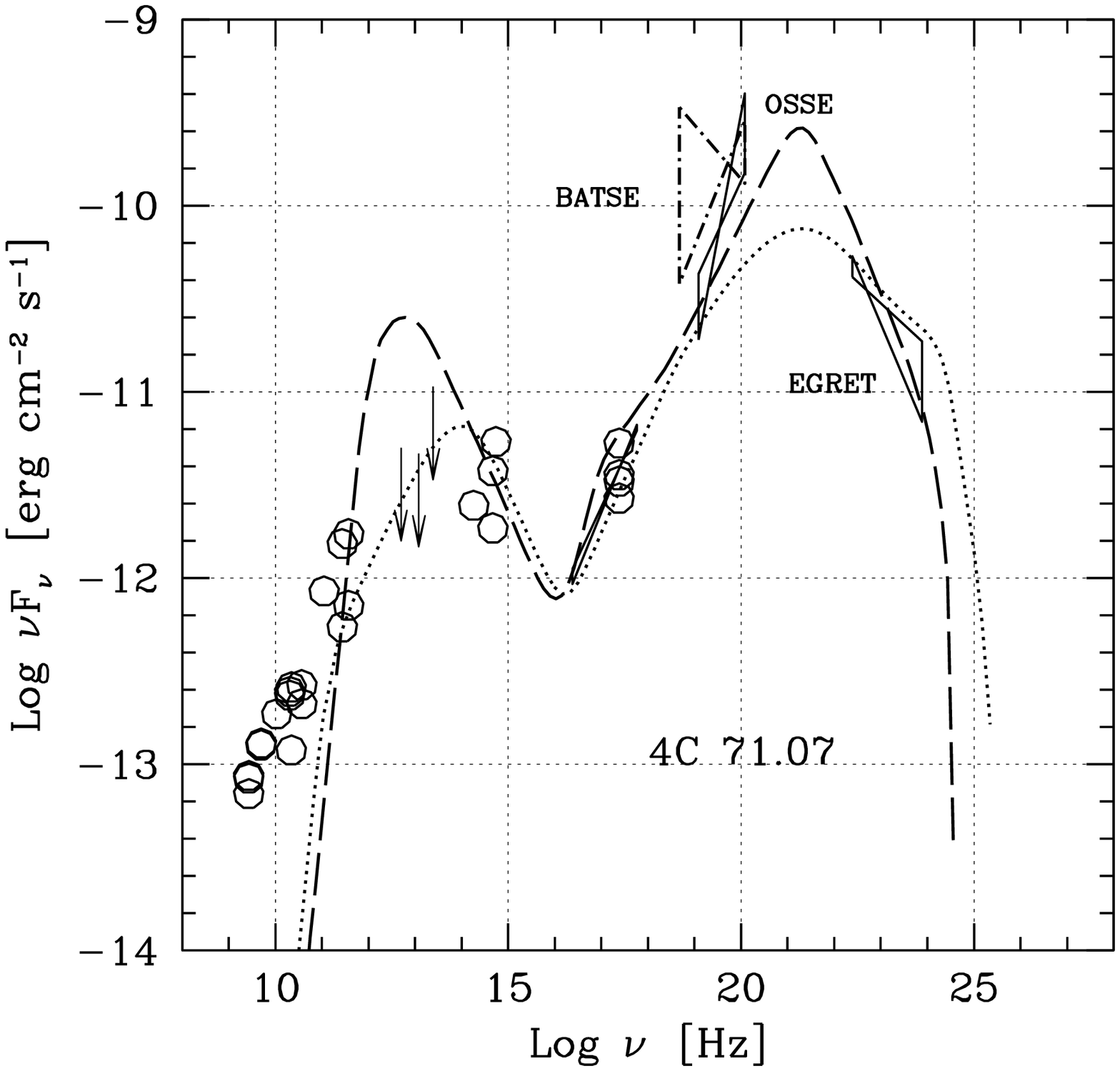]{Overall spectral energy distribution of 4C71.07
adapted from Ghisellini et al. 1998 plus BATSE and OSSE measurements
from this work and V data point from Mcintosh et al. 1999.  The 12,
25 and 60 micron upper limits have been estimated by comparison with
IRASFSC catalogue data obtained from objects located within 4 degrees
from the QSO.  The SSC and EC models are superimposed to the data as
dotted and dashed lines respectively. \label{fig4}}

%\end{document}

\clearpage

\begin{figure}
\plotone{fig1.ps}
\end{figure}

\clearpage

\begin{figure}
\plotone{fig2.ps}
\end{figure}

\clearpage

\begin{figure}
\plotone{fig3.ps}
\end{figure}

\clearpage

\begin{figure}
\plotone{fig4.ps}
\end{figure}

\end{document}